\documentclass[journal=jcp,manuscript=article]{achemso}
\setkeys{acs}{maxauthors=0}

\usepackage{achemso}
\usepackage{graphics}
\usepackage{amssymb,amsfonts}
\usepackage{graphicx}
\usepackage[table]{xcolor}
\usepackage{caption}
\usepackage{subcaption}
\usepackage{colortbl}
\usepackage{amsmath}
\usepackage{amsopn}
\usepackage{bm}
\usepackage{color}
\usepackage{array}
\usepackage{lscape}
\usepackage{mciteplus}
\usepackage[version=3]{mhchem}
\usepackage{ulem}
\usepackage{listings}
\usepackage{enumerate}
\SectionNumbersOn
\makeatletter
\newcommand*\rel@kern[1]{\kern#1\dimexpr\macc@kerna}
\newcommand*\widebar[1]{%
  \begingroup
  \def\mathaccent##1##2{%
    \rel@kern{0.8}%
    \overline{\rel@kern{-0.8}\macc@nucleus\rel@kern{0.2}}%
    \rel@kern{-0.2}%
  }%
  \macc@depth\@ne
  \let\math@bgroup\@empty \let\math@egroup\macc@set@skewchar
  \mathsurround\z@ \frozen@everymath{\mathgroup\macc@group\relax}%
  \macc@set@skewchar\relax
  \let\mathaccentV\macc@nested@a
  \macc@nested@a\relax111{#1}%
  \endgroup
}
\makeatother
\numberwithin{equation}{subsection}

\author{Janus J. Eriksen}
\email{janusje@chem.au.dk}
\affiliation[Aarhus University]
{qLEAP Center for Theoretical Chemistry, Department of Chemistry, Aarhus University, DK-8000 Aarhus C, Denmark}
\author{Devin A. Matthews}
\affiliation[The University of Texas at Austin]
{The Institute for Computational Engineering and Sciences, The University of Texas at Austin, Austin, Texas 78712, United States}
\author{Poul J\o rgensen}
\affiliation[Aarhus University]
{qLEAP Center for Theoretical Chemistry, Department of Chemistry, Aarhus University, DK-8000 Aarhus C, Denmark}
\author{J\"urgen Gauss}
\affiliation[Johannes Gutenberg-Universit\"at Mainz]
{Institut f\"ur Physikalische Chemie, Johannes Gutenberg-Universit\"at Mainz, D-55128 Mainz, Germany}

\title[TITLE]
  {Assessment of the accuracy of coupled cluster perturbation theory for open-shell systems. II. Quadruples expansions}

\begin{document}
%
%
\begin{abstract}

We extend our assessment of the potential of perturbative coupled cluster (CC) expansions for a test set of open-shell atoms and organic radicals to the description of quadruple excitations. Namely, the second- through sixth-order models of the recently proposed CCSDT(Q--$n$) quadruples series [J. Chem. Phys. {\bf{140}}, 064108 (2014)] are compared to the prominent CCSDT(Q) and $\Lambda$CCSDT(Q) models. From a comparison of the models in terms of their recovery of total CC singles, doubles, triples, and quadruples (CCSDTQ) energies, we find that the performance of the CCSDT(Q--$n$) models is independent of the reference used (unrestricted or restricted (open-shell) Hartree-Fock), in contrast to the CCSDT(Q) and $\Lambda$CCSDT(Q) models, for which the accuracy is strongly dependent on the spin of the molecular ground state. By further comparing the ability of the models to recover relative CCSDTQ total atomization energies, the discrepancy between them is found to be even more pronounced, stressing how a balanced description of both closed- and open-shell species---as found in the CCSDT(Q--$n$) models---is indeed of paramount importance if any perturbative CC model is to be of chemical relevance for high-accuracy applications. In particular, the third-order CCSDT(Q--3) model is found to offer an encouraging alternative to the existing choices of quadruples models used in modern computational thermochemistry, since the model is still only of moderate cost, albeit markedly more costly than, e.g., the CCSDT(Q) and $\Lambda$CCSDT(Q) models.

\end{abstract}
\newpage
%

%
%
\section{Introduction}\label{intro_section}

Within the field of wave function-based theoretical chemistry, the size-extensive coupled cluster~\cite{cizek_1,cizek_2,paldus_cizek_shavitt} (CC) hierarchy, which consists of the CCSD~\cite{ccsd_paper_1_jcp_1982}, CCSDT~\cite{ccsdt_paper_1_jcp_1987,*ccsdt_paper_2_cpl_1988}, CCSDTQ~\cite{ccsdtq_paper_1_jcp_1991,*ccsdtq_paper_2_jcp_1992}, etc., models with single (S), double (D), triple (T), quadruple (Q), etc., excitations, is undoubtedly the most successful family of methods for the treatment of many-body effects in atoms and molecules. While the CCSD and, to some extent, the CCSDT models have nowadays become routinely applicable, a treatment of excitation levels higher than that of triples is often mandatory, e.g., to the discipline of computational thermochemistry~\cite{ruden_helgaker_cpl_2003,w1_w2_jcp_1999,w3_jcp_2004,w4_jcp_2006,heat_1_jcp_2004,heat_2_jcp_2006,heat_3_jcp_2008,ruscic_tca_2014,ruscic_ijqc_2014} or for the accurate determination of molecular equilibrium geometries~\cite{mest,kallay_jcp_2003,szalay_jpca_2004,heckert_mol_phys_2005,heckert_jcp_2006,lane_jctc_2013} (which, in turn, is intrinsically related to the determination of accurate first- and second-order properties, harmonic frequencies, potential energy surfaces, etc.). To a first and feasible approximation, the inclusion of connected quadruple excitations proves necessary. The full iterative CCSDTQ model, however, exhibits a staggering asymptotic $\mathcal{O}(N^{10})$ scaling (where $N$ is a measure of the total system size), which in practice prevents it from routine applications for anything but atoms and small-sized molecules, although recent work has highlighted how the generally inferior convergence behavior of the model may be accelerated by performing sub-iterations of certain contributions to the higher-level CC amplitude equations~\cite{matthews_stanton_ccsdtq_jcp_2015}. For this reason, computationally tractable CC quadruples models, which exhibit reduced formal non-iterative scalings, have been devised using arguments rooted in perturbation theory. In a recent communication~\cite{quadruples_pert_theory_jcp_2015}, we performed a rigorous investigation of the potential of such non-iterative CC quadruples expansions---rationalized from either Hartree-Fock (HF) state-based many-body perturbation theory~\cite{shavitt_bartlett_cc_book} (MBPT) or CC state-based perturbation theory~\cite{ccsd_pert_theory_jcp_2014,eom_cc_pert_theory_jcp_2014}---to recover CCSDTQ total energies for a test set of {\it{closed-shell molecules}}. In summary, the study found that non-iterative CC models, in which a CCSD reference state energy is corrected for the combined effect of triple and quadruple excitations, are in general incapable of approximating the CCSDTQ model to within acceptable accuracy. As examples of such models, we tested the MBPT-rationalized CCSD(TQ$_{f}$)~\cite{kucharcki_ccsd_ptq_2_jcp_1998} and CCSD+TQ$^{\ast}$~\cite{ccsdpt_bartlett_cpl_1990} models as well as the lowest-order models of the recently proposed CCSD(TQ--$n$) perturbation series~\cite{ccsd_pert_theory_jcp_2014}, which form a hierarchy of perturbative models converging from the CCSD energy onto the CCSDTQ energy. In fact, these models, with the exception of the fourth-order CCSD(TQ--4) model, were found to exhibit absolute errors so large and erratic that they often did not even manage to improve upon the CCSD(T) model~\cite{original_ccsdpt_paper} nor the higher-level models of the CCSD(T--$n$) triples perturbation series~\cite{ccsd_pert_theory_jcp_2014,triples_pert_theory_jcp_2015}. The study thus underlined the fact that a full iterative account of triples effects---as in the CCSDT model---is integral before quadruples effects can be addressed. For this reason, the focus in the present study, which will deal with how perturbative CC quadruples expansions are capable of recovering CCSDTQ total energies, not for closed-shell molecules, but rather for {\it{open-shell species}} such as atoms and organic radicals, will be limited to variants in which the CCSDT energy is corrected for the isolated effect of quadruple excitations.\\

In the first part of the present series~\cite{open_shell_triples_arxiv_2015}, we investigated the accuracy at which total energies of open-shell species are determined using CC perturbative triples expansions. In this second part, we wish to extend the analysis to CC perturbative quadruples expansions, with special attention given to the models of the recently proposed CCSDT(Q--$n$) series, which---like the CCSD(TQ--$n$) series---forms a hierarchy of perturbative models, but now converging from the CCSDT energy, rather than the CCSD energy, onto the CCSDTQ energy~\cite{quadruples_pert_theory_jcp_2015,ccsd_pert_theory_jcp_2014}. More precisely, the CCSDT(Q--$n$) series is defined as an order expansion in the M{\o}ller-Plesset fluctuation potential of a bivariational CCSDTQ energy Lagrangian where, from a CCSDT zeroth-order expansion point, perturbative solutions of both the exponential CCSDTQ amplitude equations and linear $\Lambda$-state equations are embedded into the energy corrections~\cite{e_ccsd_tn_jcp_2016}. Results will be reported for the second- through sixth-order models of the CCSDT(Q--$n$) series; for the fifth- and sixth-order models---the CCSDT(Q--5) and CCSDT(Q--6) models, respectively---results for both closed- and open-shell systems are presented for the first time. Furthermore, we will relate the potential of a given perturbative model to recover total CCSDTQ energies to its potential as a tool in computational thermochemistry, with an initial application to simple total atomization energies (TAEs). In this respect, we note how all of thermochemistry is concerned with appropriately defined energy differences rather than total energies of molecules. Thus, a prerequisite for the application of any given model is not necessarily to provide results of thermochemical accuracy (i.e., within the sub-kJ/mol range) for total energies, but rather to achieve this level of accuracy for relative energy differences. 

In Ref. \citenum{quadruples_pert_theory_jcp_2015}, TAEs were calculated for a test set of small-sized closed-shell molecules. However, transient species, e.g., open-shell atoms and radicals, are omnipresent in some of the fields that rely most strongly on accurate thermochemical data, such as combustion and atmospheric chemistry, and while most closed-shell molecules are amenable to experimental characterization, this is not the case for open-shell species, for which the experimental error bars are typically significantly larger~\cite{heat_1_jcp_2004}. For this reason, high-level {\it{ab initio}} computational chemistry has a significant role to play in the field of thermochemistry, not only as a supplement to experimental laboratory work, but also as a generally practical and viable approach in the case of transient species. As noted previously in Ref. \citenum{heat_2_jcp_2006}, calculations of total energies for various closed- and open-shell species may potentially benefit from cancellations of different types of errors. However, for any theoretical method to be truly reliable, its errors---both in terms of sources and magnitude---have to be balanced between calculations on closed- and open-shell species. Phrased differently, if pronounced and unsystematic discrepancies exist for a given model in its description of total energies for closed- and/or open-shell species, said model will owe its performance (good or bad) for relative quantities, such as thermochemical parameters, as much to fortuitous and unpredictable error cancellations as to theoretical rigor.\\

In the present work, we will compare the performance of the CCSDT(Q--$n$) models for both total and atomization energies (calculated in modest-sized basis sets of double-$\zeta$ quality) to the CCSDT(Q)~\cite{bomble_ccsdt_pq_jcp_2005} and $\Lambda$CCSDT(Q)~\cite{kallay_gauss_2005} models. However, since we want to report open-shell results for both unrestricted HF (UHF) and restricted open-shell HF (ROHF) trial functions~\cite{pople_nesbet_uhf_jcp_1954}, we note that the generalization of the $\Lambda$CCSDT(Q) model to both sets of references is unambiguous, unlike for the CCSDT(Q) model, for which, in the case of an ROHF reference, two variants have been proposed in the literature~\cite{kallay_gauss_2008}. For this reason, we will only compare the CCSDT(Q--$n$) models to the CCSDT(Q) model for UHF references. 

%
%
\section{Computational details}\label{com_details_section}

All of the perturbative CC quadruples expansions tested here are based on the CCSDT model, either for a UHF or a (semicanonical) ROHF reference~\cite{watts_bartlett_rohf_ccsdt_jcp_1990}, and compared against the CCSDTQ model for a general single-reference determinant~\cite{olsen_general_order_cc_jcp_2000,kallay_string_based_cc_jcp_2001}. While the CCSDTQ model scales iteratively as $\mathcal{O}(N^{10})$, the formal scaling of the CCSDT(Q) and $\Lambda$CCSDT(Q) models is only $\mathcal{O}(N^{9})$, although for the latter of the models, both the CCSDT amplitude and $\Lambda$-state equations~\cite{handy_schaefer_lambda_ci_jcp_1984,*schaefer_lambda_cc_jcp_1987} have to be converged prior to the evaluation of the actual quadruples correction. All of the CCSDT(Q--$n$) models, on the other hand, scale non-iteratively as $\mathcal{O}(N^{10})$, with an increasing number of cost-determining contractions appearing at higher orders upon moving up through the CCSDT(Q--$n$) series. For example, for the difficult closed-shell ozone molecule, the increase in compute time over the CCSDT solution was previously found to be approximately a factor of 2 (4) and 4 (7) for the CCSDT(Q) and $\Lambda$CCSDT(Q) models, respectively, in a cc-pVDZ (cc-pVTZ) basis~\cite{dunning_1_orig,*dunning_5_core}, with obvious larger factors for the CCSDT(Q--$n$) models, cf. the cost discussion in Ref. \citenum{quadruples_pert_theory_jcp_2015}. However, comparing the cost of the CCSDT(Q--3) and CCSDT(Q--4) models, for instance, to that of the $\Lambda$CCSDT(Q) model, revealed how the total time-to-solution was only increased by a factor of 5 and 10, respectively, for the evaluation of the former two corrections over the latter in a cc-pVTZ basis, and less so in the smaller cc-pVDZ basis~\cite{quadruples_pert_theory_jcp_2015}.\\

The numerical performance of each of the models is reported on par with Ref. \citenum{open_shell_triples_arxiv_2015}, namely in terms of {\textbf{(i)}} the relative recovery of the contribution to the CCSDTQ correlation energy from quadruple excitations (i.e., the CCSDTQ--CCSDT correlation energy difference) and {\textbf{(ii)}} the actual deviation from this difference. For the study of TAEs, we report the total deviations from the CCSDTQ results, and for individual closed- and open-shell results, we refer to the supplementary material~\bibnote{See supplementary material at [AIP URL] for individual recoveries and deviations.}. All of the CCSDT(Q) and $\Lambda$CCSDT(Q) calculations have been performed within the general string-based CC code~\cite{kallay_string_based_cc_jcp_2001,kallay_gauss_2005,kallay_gauss_2008} of the \textsc{mrcc} program~\cite{mrcc}, to which an interface from the \textsc{cfour} quantum chemical program package~\cite{cfour} exists, while the recently developed Aquarius program~\cite{aquarius} has been used for all of the CCSDT(Q--$n$) calculations~\bibnote{The comments made in Ref. \citenum{open_shell_triples_arxiv_2015} concerning the ROHF-based CCSD(T--$n$) implementations in Aquarius apply directly to the ROHF-based CCSDT(Q--$n$) implementations of the present work.}. The closed-shell~\bibnote{Closed-shell test set: H$_2$O; H$_2$O$_2$; CO; CO$_2$; C$_2$H$_2$; C$_2$H$_4$; CH$_2$ ($^{1}\text{A}_{1}$); CH$_2$O; N$_2$; NH$_3$; N$_2$H$_2$; HCN; HOF; HNO; F$_2$; HF; O$_3$. Geometries are listed in Ref. \citenum{mest}.} and open-shell~\bibnote{Open-shell test set: C; CCH; CF; CH; CH$_{2}$ ($^3\text{B}_1$); CH$_3$; CN; F; HCO; HO$_2$; N; NH; NH$_2$; NO; O; O$_2$; OF; OH. Geometries are listed in Ref. \citenum{heat_1_jcp_2004}.} test sets are those listed in Ref. \citenum{open_shell_triples_arxiv_2015}, in which a discussion on the amount of spin contamination remaining at the CCSDT and CCSDTQ levels of theory may also be found, and for all of the reported valence-electron (frozen-core) results of the present work, the correlation-consistent cc-pVDZ basis set has been used.

%
%
\section{Results}\label{results_section}
\begin{figure}
        \centering
        \begin{subfigure}[b]{0.47\textwidth}
                \includegraphics[width=\textwidth,bb=0 0 488 384]{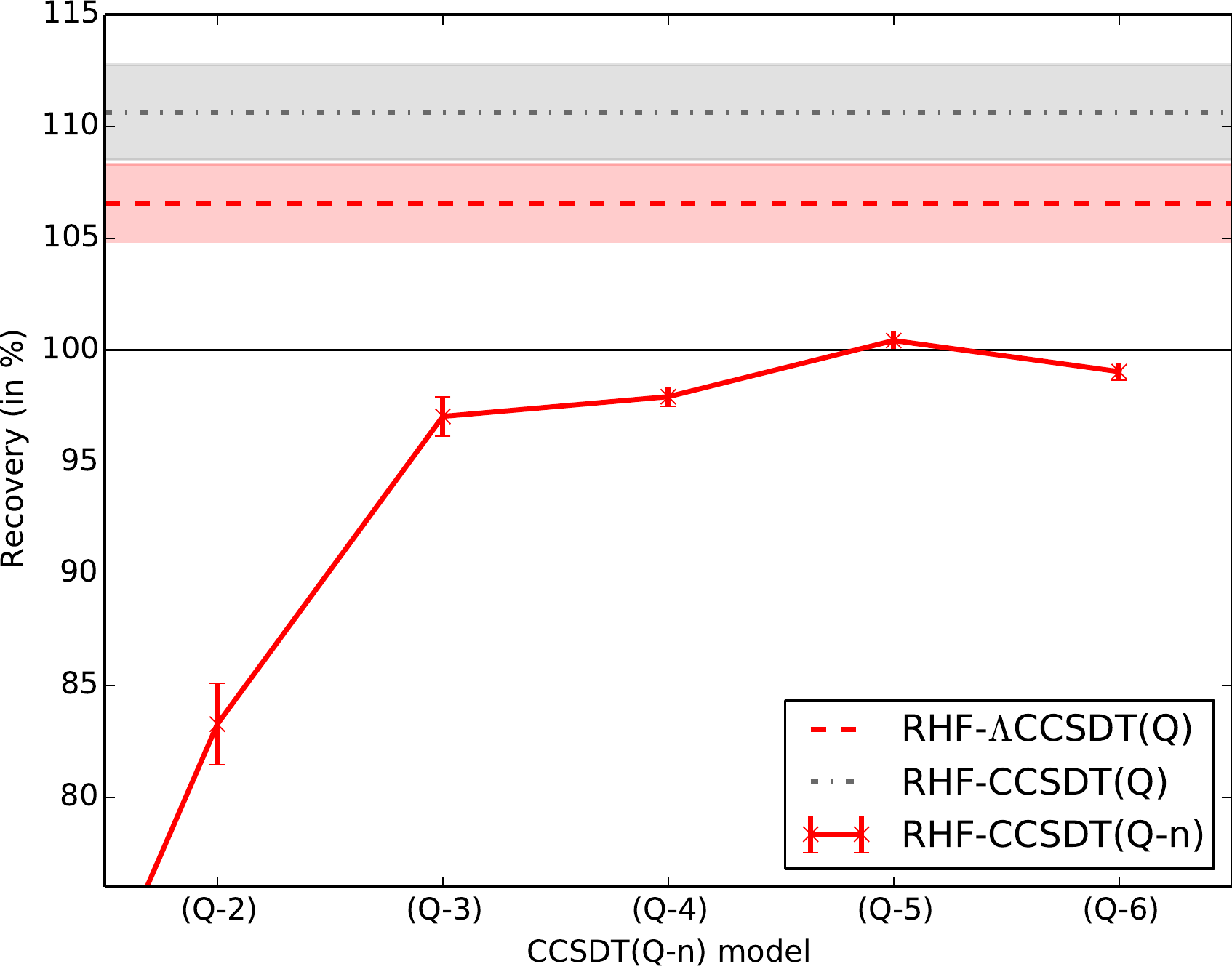}
                \caption{RHF recoveries}
                \label{q_n_recoveries_rhf_figure}
        \end{subfigure}%
        ~ 
        \begin{subfigure}[b]{0.47\textwidth}
                \includegraphics[width=\textwidth,bb=0 0 488 384]{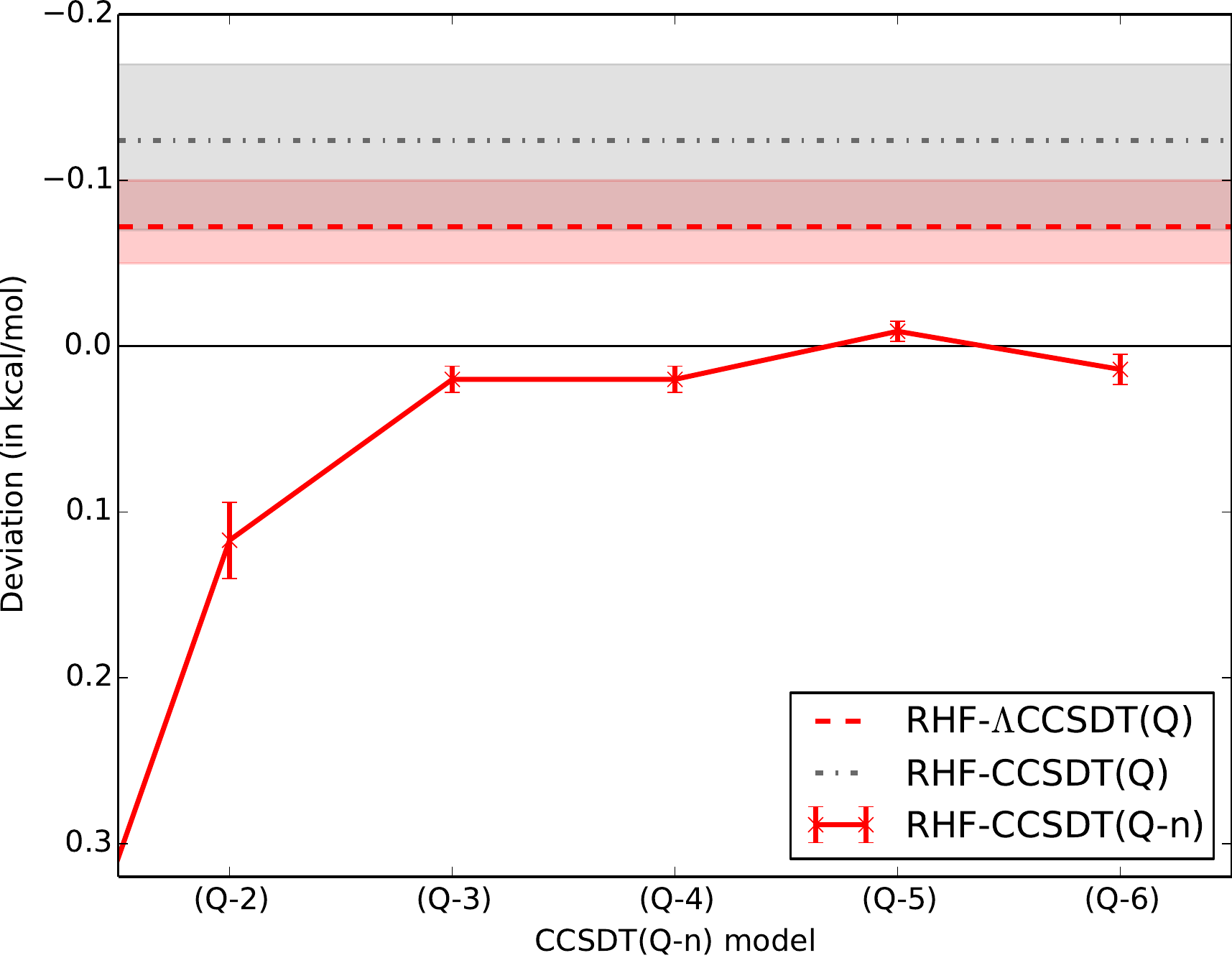}
                \caption{RHF deviations}
                \label{q_n_abs_diff_rhf_figure}
        \end{subfigure}
        \begin{subfigure}[b]{0.47\textwidth}
                \includegraphics[width=\textwidth,bb=0 0 488 384]{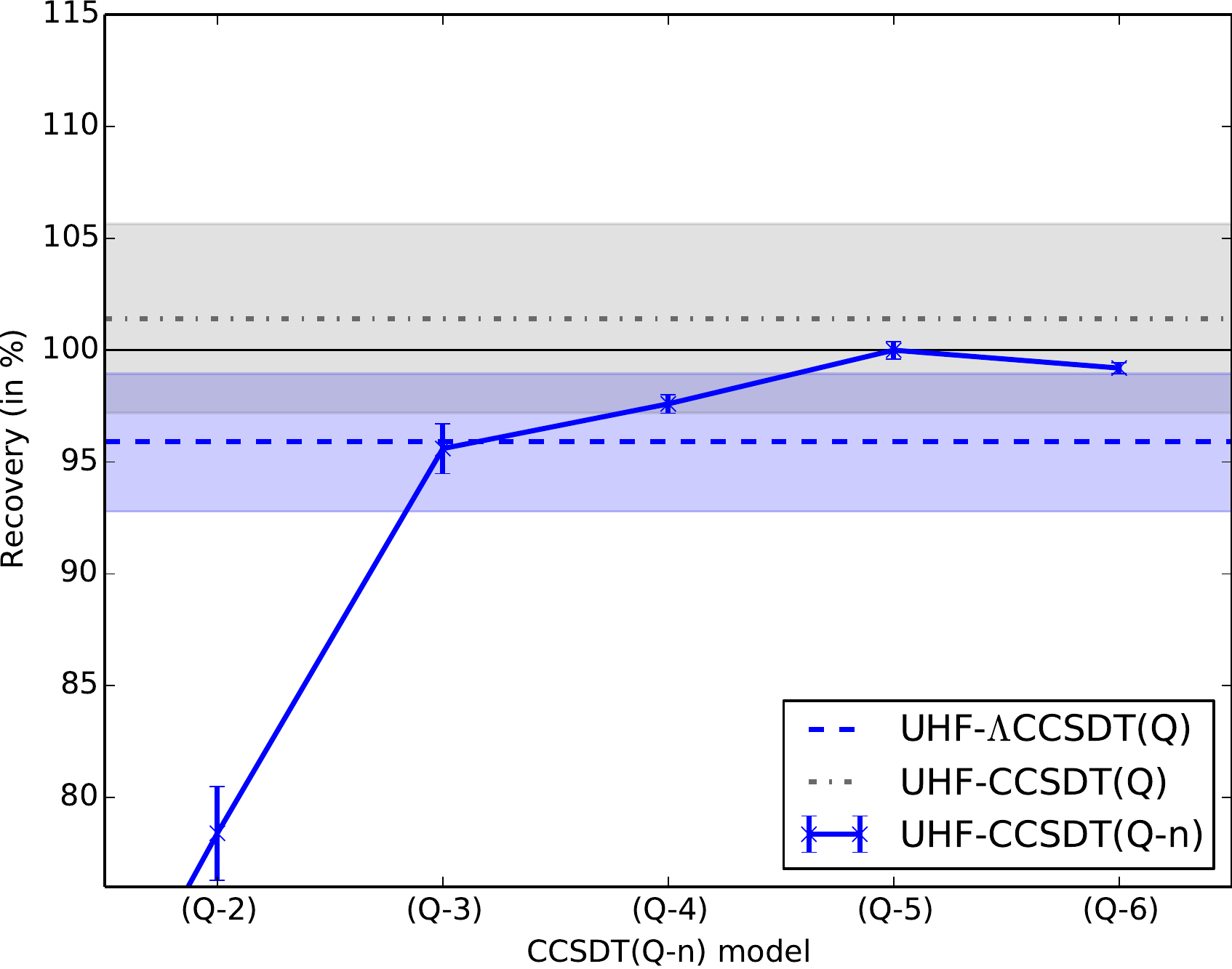}
                \caption{UHF recoveries}
                \label{q_n_recoveries_uhf_figure}
        \end{subfigure}%
        ~ 
        \begin{subfigure}[b]{0.47\textwidth}
                \includegraphics[width=\textwidth,bb=0 0 488 384]{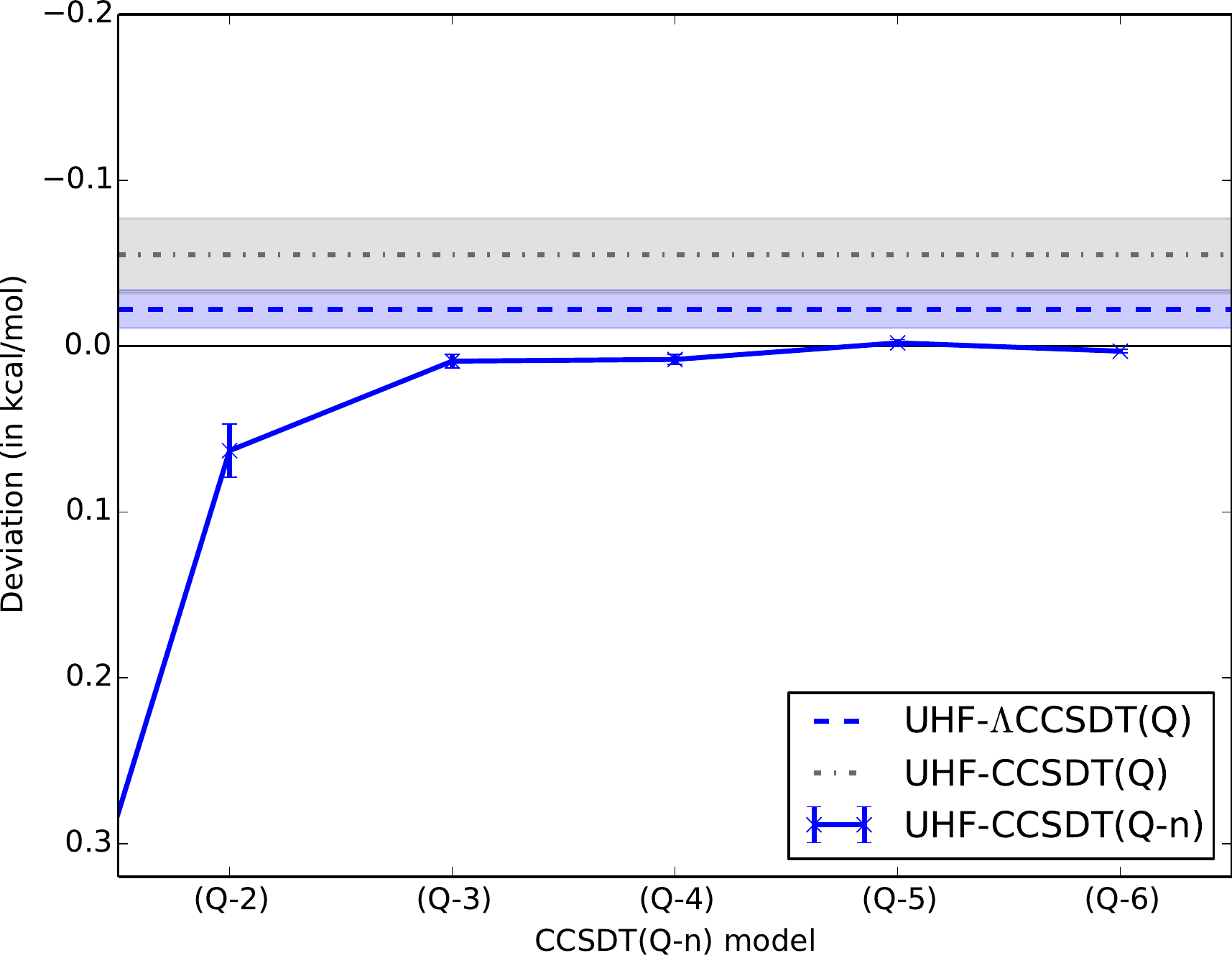}
                \caption{UHF deviations}
                \label{q_n_abs_diff_uhf_figure}
        \end{subfigure}
        \begin{subfigure}[b]{0.47\textwidth}
                \includegraphics[width=\textwidth,bb=0 0 488 384]{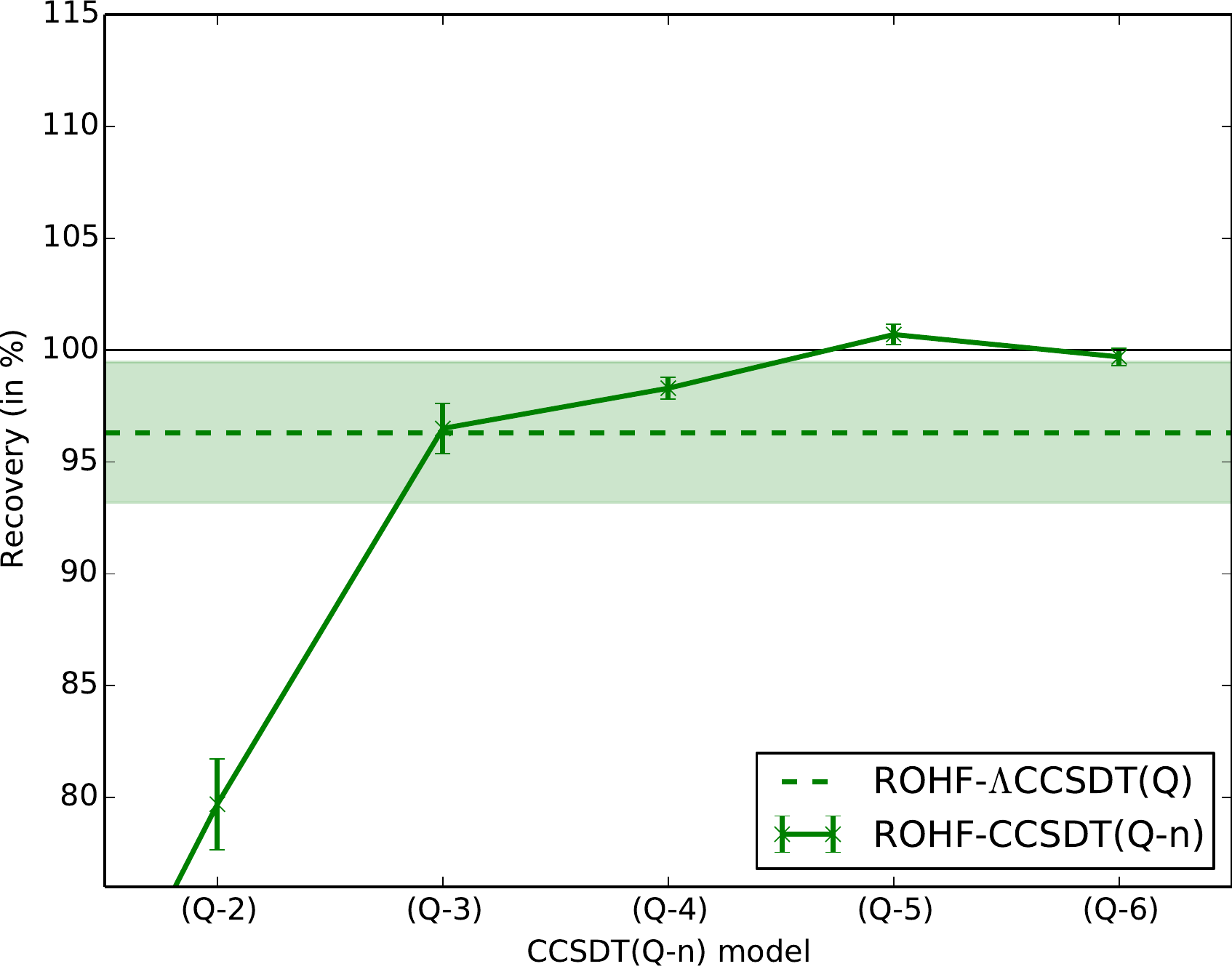}
                \caption{ROHF recoveries}
                \label{q_n_recoveries_rohf_figure}
        \end{subfigure}%
        ~ 
        \begin{subfigure}[b]{0.47\textwidth}
                \includegraphics[width=\textwidth,bb=0 0 488 384]{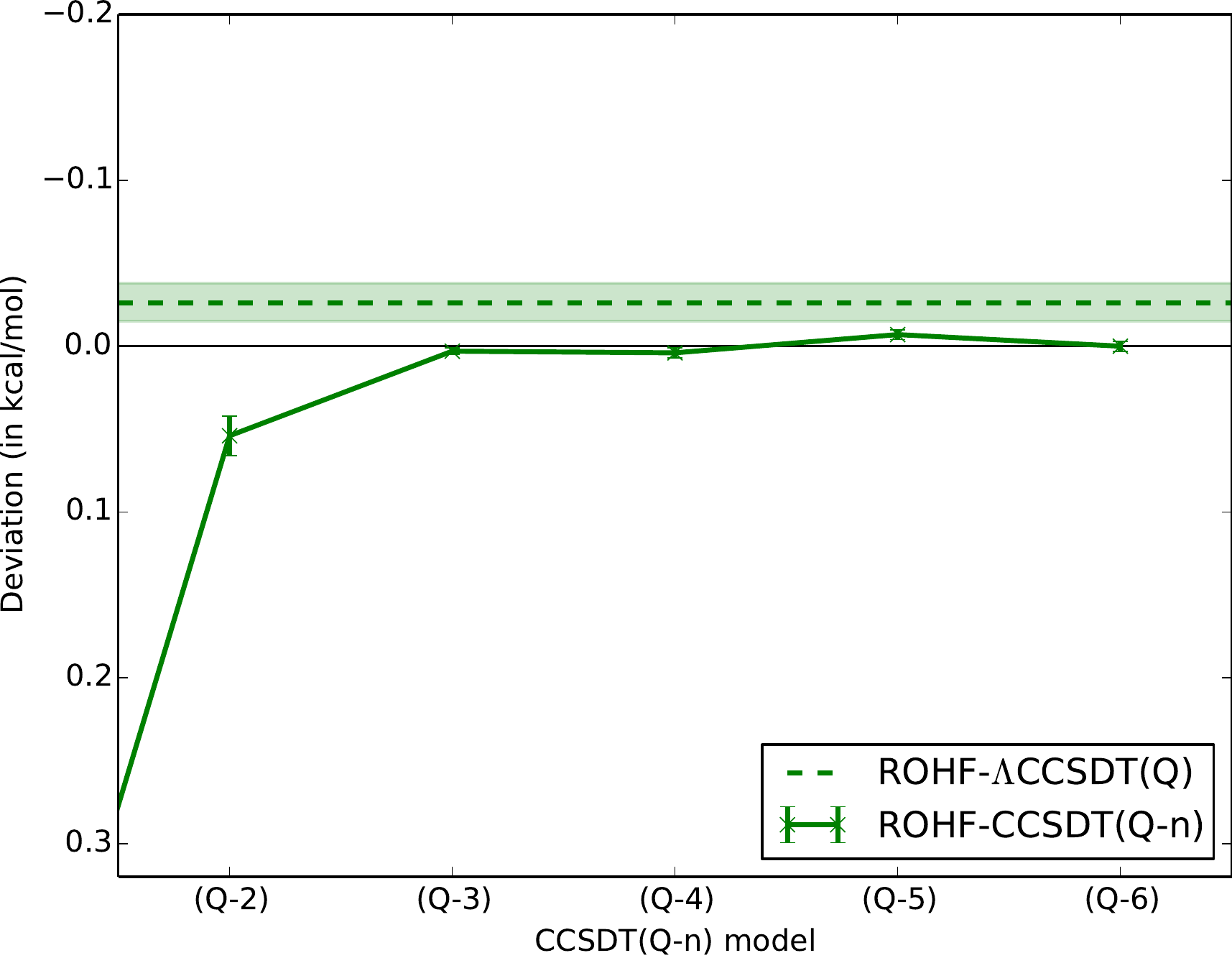}
                \caption{ROHF deviations}
                \label{q_n_abs_diff_rohf_figure}
        \end{subfigure}
        \caption{Recoveries of (in percent (\%), Figures \ref{q_n_recoveries_rhf_figure}, \ref{q_n_recoveries_uhf_figure}, and \ref{q_n_recoveries_rohf_figure}) and deviations from (in kcal/mol, Figures \ref{q_n_abs_diff_rhf_figure}, \ref{q_n_abs_diff_uhf_figure}, and \ref{q_n_abs_diff_rohf_figure}) CCSDTQ--CCSDT frozen-core/cc-pVDZ correlation energy differences for RHF, UHF, and ROHF references. The error bars show the standard error of the mean, and for the CCSDT(Q) and $\Lambda$CCSDT(Q) models, the standard errors are depicted as colored intervals centred around the mean.}
        \label{q_n_figure}
\end{figure}

\begin{figure}[H]
        \centering
        \begin{subfigure}[b]{0.47\textwidth}
                \includegraphics[width=\textwidth,bb=0 0 488 384]{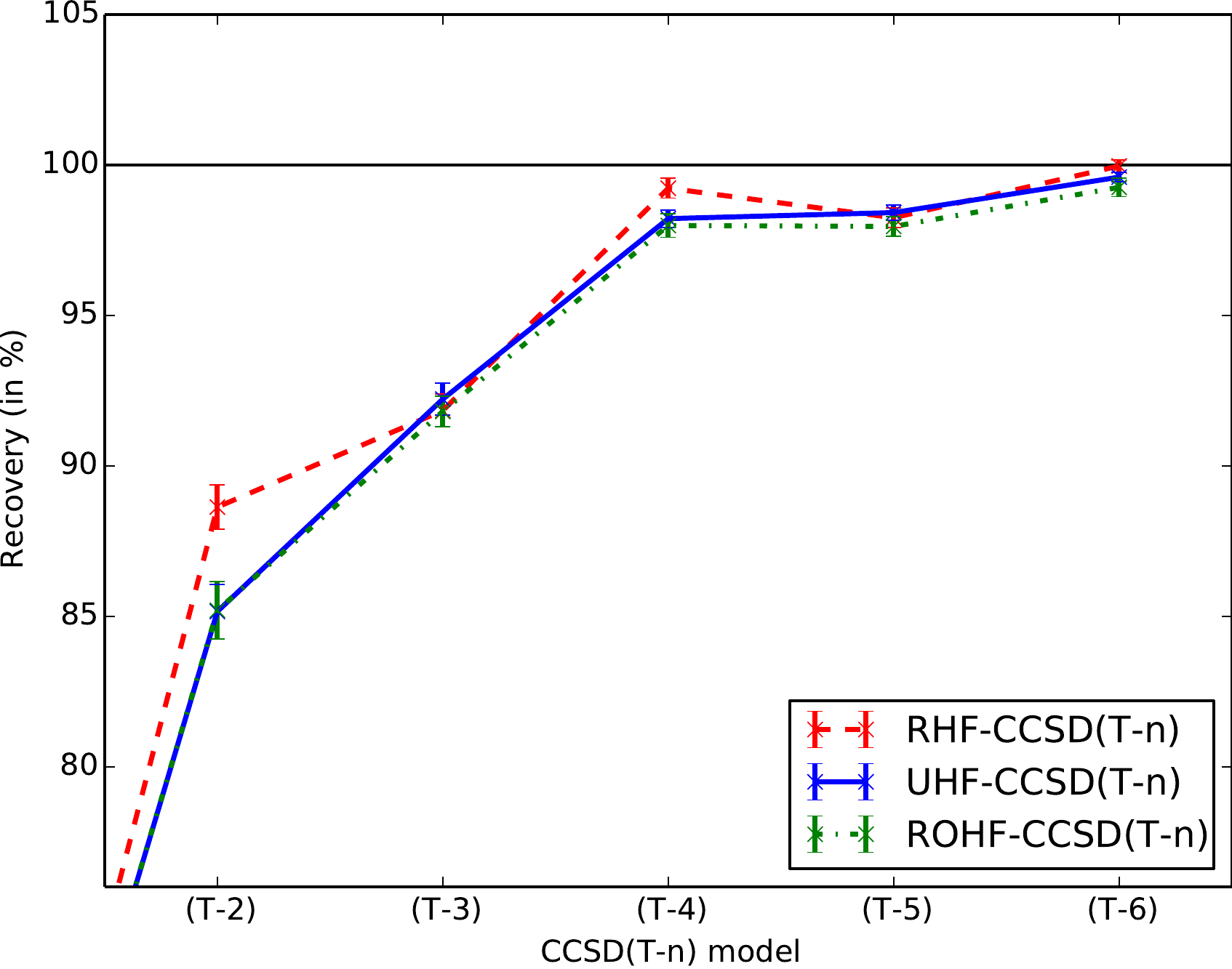}
                \caption{CCSD(T--$n$) recoveries}
                \label{t_n_comparison_rec_figure}
        \end{subfigure}%
        \hspace{0.4cm}
        ~
        \begin{subfigure}[b]{0.47\textwidth}
                \includegraphics[width=\textwidth,bb=0 0 488 384]{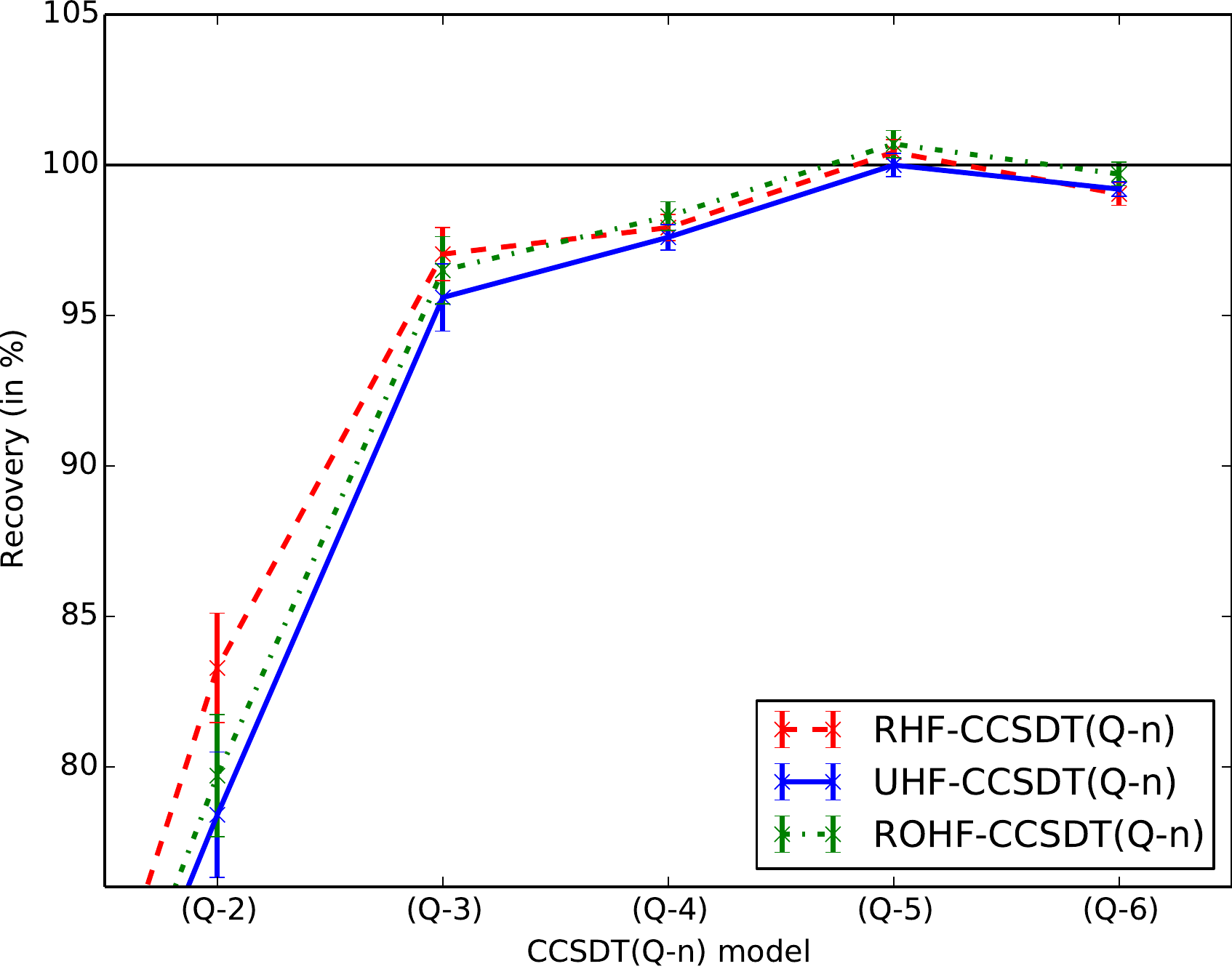}
                \caption{CCSDT(Q--$n$) recoveries}
                \label{q_n_comparison_rec_figure}
        \end{subfigure}
        \begin{subfigure}[b]{0.47\textwidth}
                \includegraphics[width=\textwidth,bb=0 0 488 384]{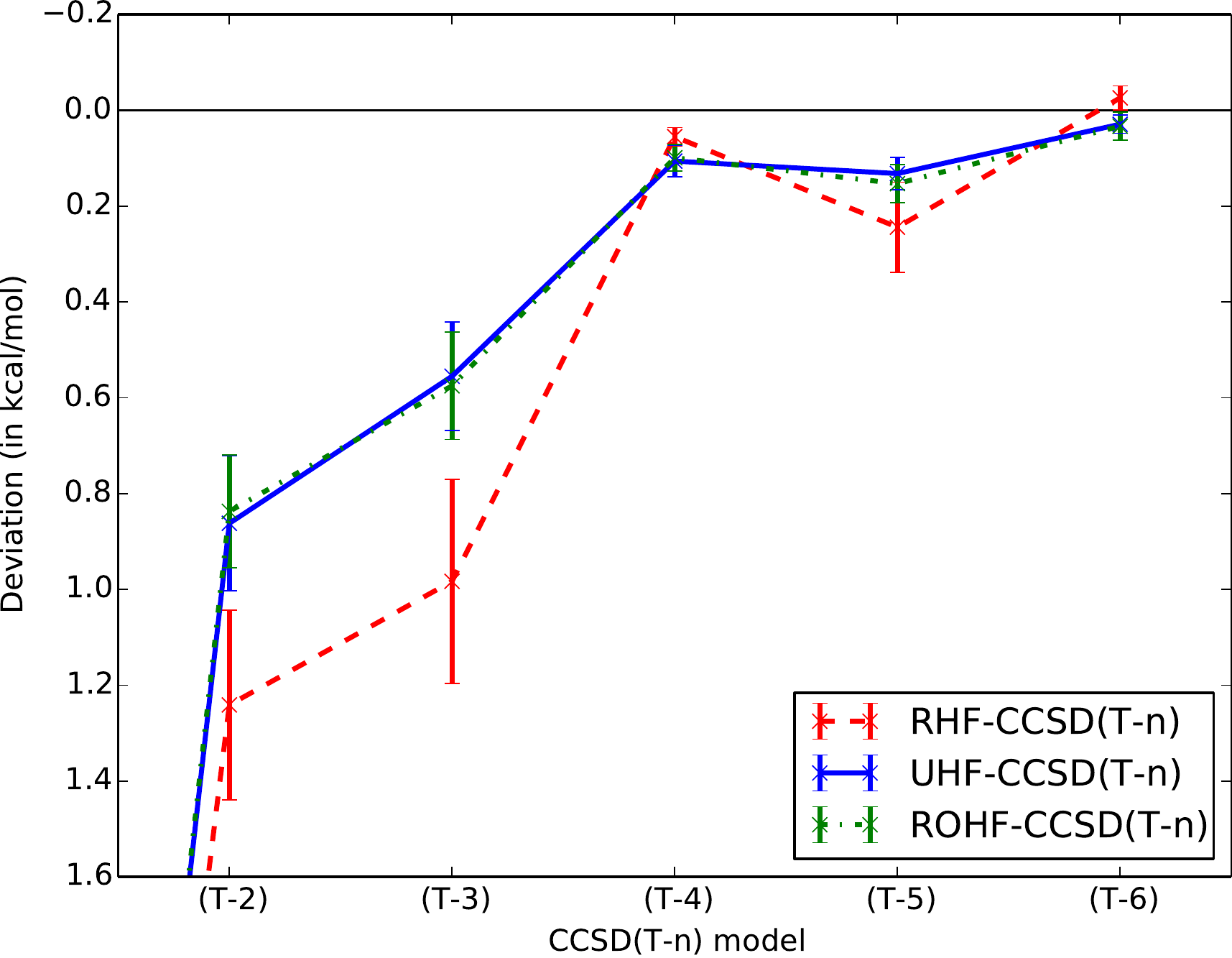}
                \caption{CCSD(T--$n$) deviations}
                \label{t_n_comparison_diff_figure}
        \end{subfigure}%
        \hspace{0.4cm}
        ~
        \begin{subfigure}[b]{0.47\textwidth}
                \includegraphics[width=\textwidth,bb=0 0 488 384]{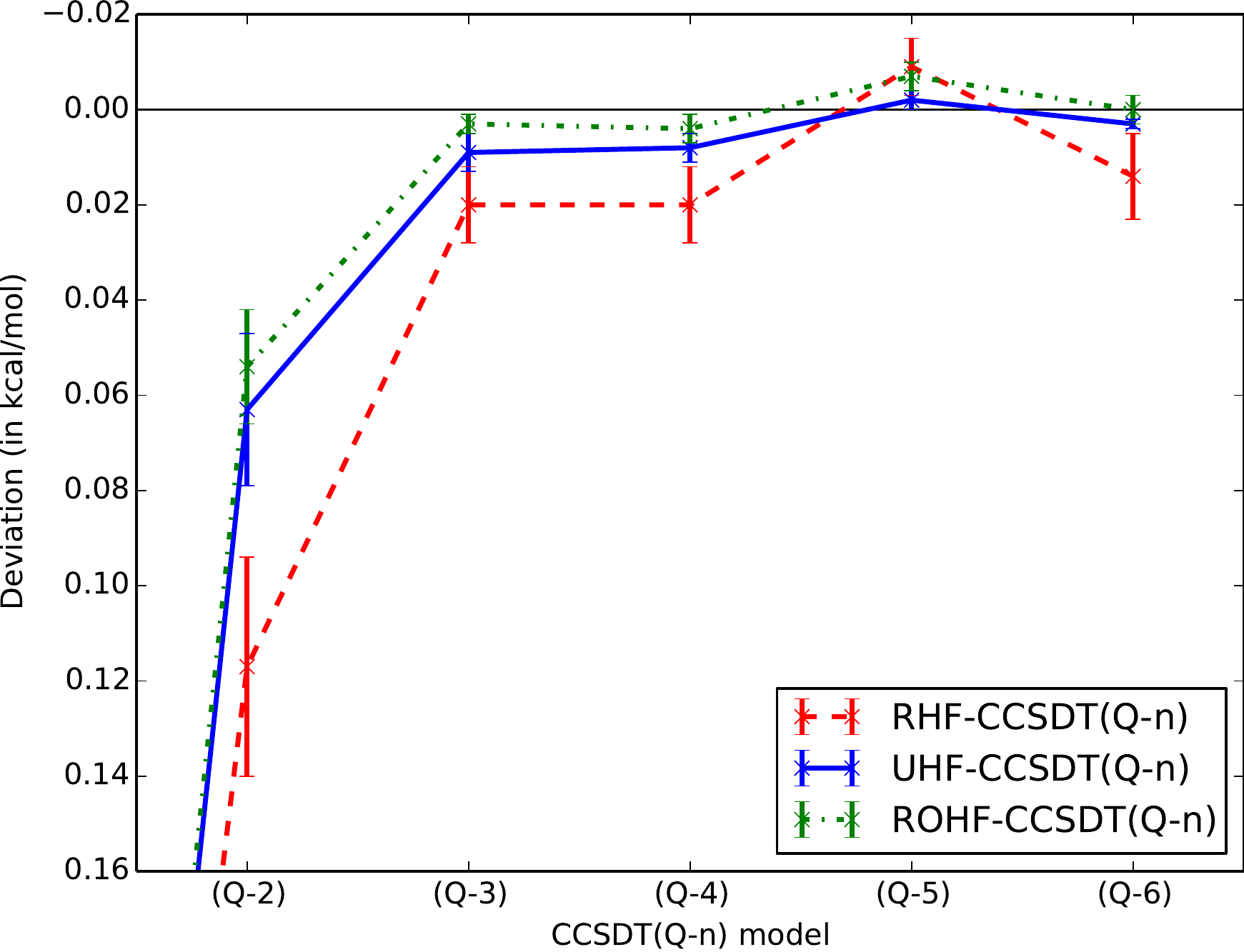}
                \caption{CCSDT(Q--$n$) deviations}
                \label{q_n_comparison_diff_figure}
        \end{subfigure}
   \caption{Recoveries of (in percent (\%)) and deviations from (in kcal/mol) CCSDT--CCSD frozen-core/cc-pVTZ and CCSDTQ--CCSDT frozen-core/cc-pVDZ correlation energy differences for the CCSD(T--$n$) (Figures \ref{t_n_comparison_rec_figure} and \ref{t_n_comparison_diff_figure}) and CCSDT(Q--$n$) (Figures \ref{q_n_comparison_rec_figure} and \ref{q_n_comparison_diff_figure}) series, respectively. In all figures, results for RHF, UHF, and ROHF references are compared next to one another.}
   \label{comparison_figure}
\end{figure}

In Figure \ref{q_n_figure}, UHF- and ROHF-based results are presented for all of the tested models, again, as in Ref. \citenum{open_shell_triples_arxiv_2015}, with RHF-based results included for a full comparison (for the CCSDT(Q) model, only RHF- and UHF-based results are reported, cf. the closing paragraph of Section \ref{intro_section}). Inspecting first the recovery of the CCSDTQ quadruples contribution in Figures \ref{q_n_recoveries_rhf_figure}, \ref{q_n_recoveries_uhf_figure}, and \ref{q_n_recoveries_rohf_figure}, we note how the theoretically predicted convergence of the CCSDT(Q--$n$) series towards the CCSDTQ target energy is numerically confirmed through sixth order in the perturbation, and, most importantly, how this holds true not only for closed-shell species, but regardless of the spin of the ground state. Furthermore, the three different curves for RHF, UHF, and ROHF are remarkably similar, even more so than what was previously observed for the CCSD(T--$n$) series in Ref. \citenum{open_shell_triples_arxiv_2015}. In fact, it is instructive to compare the results for the CCSD(T--$n$) and CCSDT(Q--$n$) series {\it{vis-\`a-vis}}. This is done in Figure \ref{comparison_figure}, in which size-intensive recoveries as well as total deviations for all three HF references are compared next to one another for both series (for the CCSD(T--$n$) series, recoveries and deviations are plotted with respect to the triples contribution to the CCSDT correlation energy). As is obvious from comparing the CCSD(T--$n$) series in Figures \ref{t_n_comparison_rec_figure} and \ref{t_n_comparison_diff_figure} with the CCSDT(Q--$n$) series in Figures \ref{q_n_comparison_rec_figure} and \ref{q_n_comparison_diff_figure}, the latter converges at a slightly faster rate than the former, most likely due to the smaller magnitude of the contribution from quadruple excitations in the CCSDTQ model than the contribution from triple excitations in the CCSDT model (note the different axes used in Figures \ref{t_n_comparison_diff_figure} and \ref{q_n_comparison_diff_figure}). Upon moving up through either of the two hierarchies, however, an identical behavior is observed, for recoveries as well as total deviations; for instance, at the fourth-order level (the CCSD(T--4) and CCSDT(Q--4) models), a relative error against either of the target energies (CCSDT and CCSDTQ) of less than $2.5$ percentage points is observed for all HF reference functions (Figures \ref{t_n_comparison_rec_figure} and \ref{q_n_comparison_rec_figure}, respectively). The error then diminishes for both of the series upon moving to higher orders, and at the sixth-order level (the CCSD(T--6) and CCSDT(Q--6) models), both series are practically converged onto their respective target energy.

Returning to Figure \ref{q_n_figure}, we note how the CCSDT(Q) results are generally in worse agreement with the CCSDTQ target energy than the $\Lambda$CCSDT(Q) results, which, in turn, are significantly worse than those of any of the CCSDT(Q--$n$) models, with the exception of the lowest-order CCSDT(Q--2) model. While in terms of the relative recoveries in Figures \ref{q_n_recoveries_rhf_figure}, \ref{q_n_recoveries_uhf_figure}, and \ref{q_n_recoveries_rohf_figure}, the performance of the CCSDT(Q) and $\Lambda$CCSDT(Q) models does not appear too poor, things change for the worse upon evaluating their performance in terms of total deviations from the CCSDTQ correlation energies. For instance, the standard deviations of the models, in particular for the closed-shell, but also for the open-shell test set, are unacceptably large. For the closed-shell results in Figure \ref{q_n_abs_diff_rhf_figure}, the statistical results for all of the models (also those of the CCSDT(Q--$n$) series) are significantly tainted by the individual results for the ozone molecule, cf. Tables S1 and S4 of the supplementary material. If one were to omit the contribution from the O$_3$ results to the results in Figures \ref{q_n_recoveries_rhf_figure} and \ref{q_n_abs_diff_rhf_figure}, these would be much improved for all of the models, to such an extent that the discrepancy between the performance of the CCSDT(Q) and $\Lambda$CCSDT(Q) models for closed- and open-shell systems would become reminiscent of the performance observed for the CCSD(T) triples model in the first part of the present series~\cite{open_shell_triples_arxiv_2015}, in spite of the fact that the $\Lambda$CCSDT(Q) model constitutes a clear improvement over the CCSDT(Q) model for both test sets.

\begin{figure}
        \centering
        \begin{subfigure}[b]{0.47\textwidth}
                \includegraphics[width=\textwidth,bb=0 0 495 384]{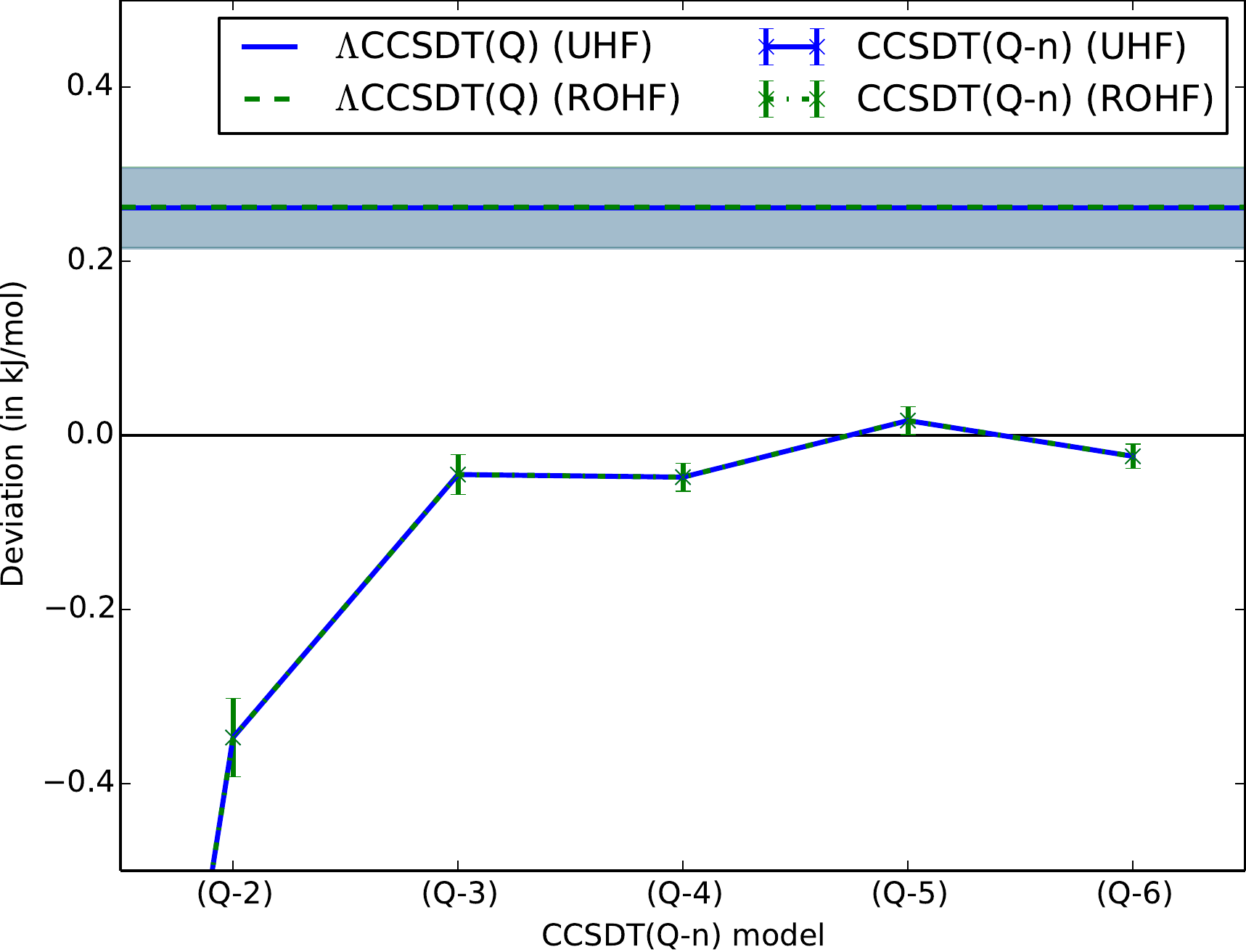}
                \caption{Closed-shell}
                \label{ae_closed_figure}
        \end{subfigure}%
        \hspace{0.4cm} 
        \begin{subfigure}[b]{0.47\textwidth}
                \includegraphics[width=\textwidth,bb=0 0 495 384]{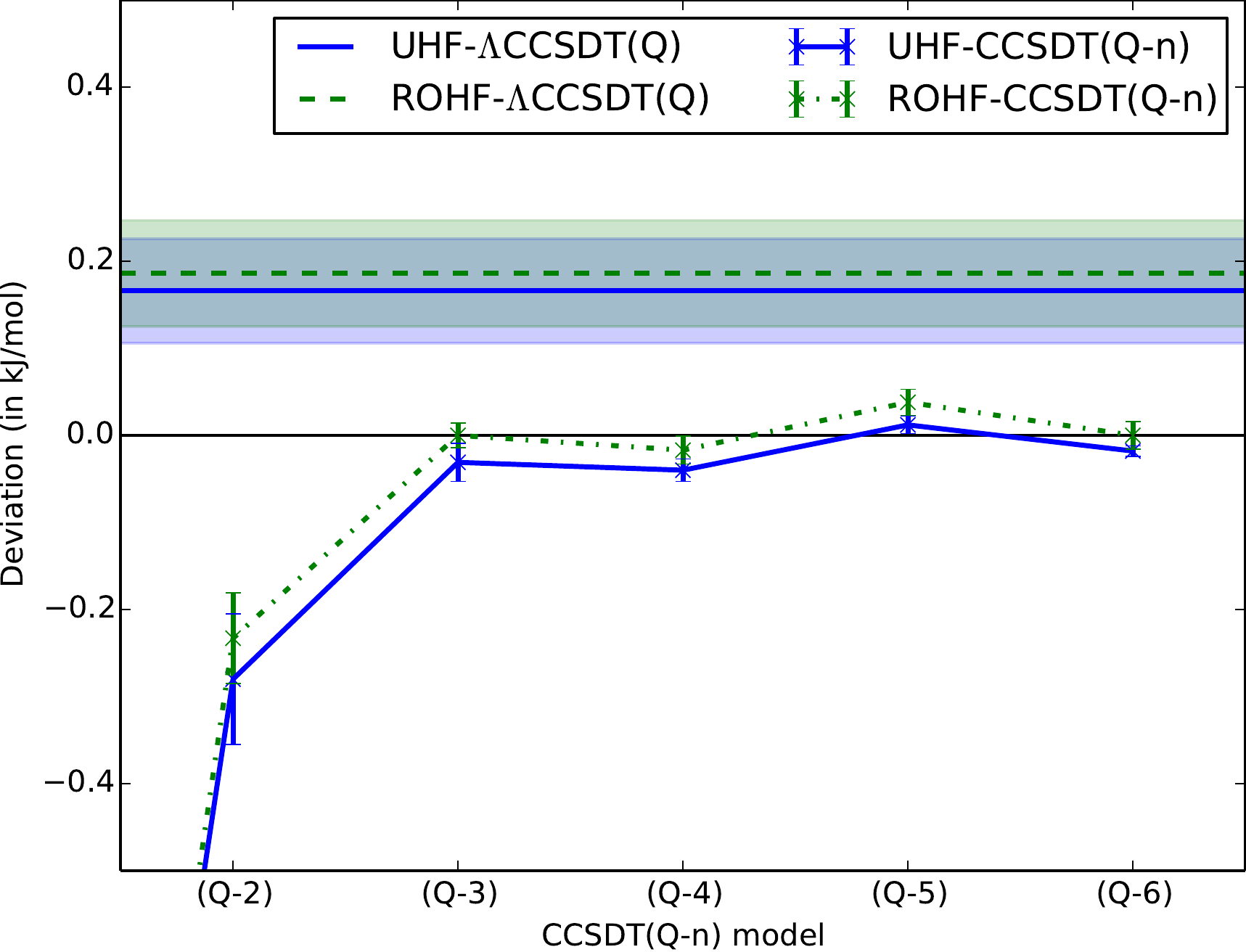}
                \caption{Open-shell}
                \label{ae_open_figure}
        \end{subfigure}
   \caption{Total deviations (in kJ/mol) from CCSDTQ frozen-core/cc-pVDZ TAEs for the two test sets of closed-shell (Figure \ref{ae_closed_figure}, with O$_3$ omitted) and open-shell (Figure \ref{ae_open_figure}) species. In both figures, the type of open-shell reference used has been indicated (in Figure \ref{ae_closed_figure}, the UHF and ROHF labels designate the HF reference used in the atomic calculations). The error bars show the standard error of the mean, and for the $\Lambda$CCSDT(Q) model, the standard error is depicted as in Figure \ref{q_n_figure}.}
   \label{ae_figure}
\end{figure}
As mentioned in Section \ref{intro_section}, for any approximate method---be it non-iterative or iterative in the computationally cost-determining step---to be of chemical relevance for high-accuracy applications, it must achieve an acceptable level of accuracy for relative energy differences. As such, it matters not how accurate the given method is in calculating total electronic energies, since there might exist a number of common sources for potential disagreements with higher-level references, which might, in turn, lead to beneficial error cancellations. However, for such cancellations of errors to be successful, the errors will not only have to be of similar magnitude, as they must also be of equal sign in-between calculations. For instance, a discrepancy in the accuracy (against a well-defined yardstick, such as the CCSDTQ model) with which a method calculates total energies for closed- and open-shell species will ultimately result in unpredictable and unsystematic errors for relative calculations that involve both types of species. In other words, a given method will have to be balanced for a variety of species, which might differ in spin, charge, molecular distortion, or the energetic gap between the ground state and possible low-lying excited states (i.e., degeneracy effects). Thus, in the present context, one cannot discard, e.g., the $\Lambda$CCSDT(Q) model (or {\it{vice versa}} favor the higher-level CCSDT(Q--$n$) models) exclusively on the basis of results for total energies like the ones presented in Figure \ref{q_n_figure}. For this reason, we present (in Figure \ref{ae_figure}) $\Lambda$CCSDT(Q) and CCSDT(Q--$n$) TAEs, based on the results in Figure \ref{q_n_figure}, i.e., for the same two test sets (O$_3$ omitted from the closed-shell test set), and reported in terms of the total deviation from results obtained with the CCSDTQ model (for both the $\Lambda$CCSDT(Q) and CCSDT(Q--$n$) results in Figure \ref{ae_closed_figure}, the UHF- and ROHF-based results (UHF or ROHF references used in the atomic calculations) are similar to such an extent that they are practically indistinguishable). As is clear---despite generally performing better than the CCSDT(Q) model---the errors of the $\Lambda$CCSDT(Q) model are substantial, even more so than for the total energies in Figures \ref{q_n_abs_diff_rhf_figure}, \ref{q_n_abs_diff_uhf_figure}, and \ref{q_n_abs_diff_rohf_figure}. Furthermore, contrary to the results in Figure \ref{q_n_figure}, the $\Lambda$CCSDT(Q) results for the closed-shell test set are now in worse agreement with the CCSDTQ reference energies than those for the open-shell test set (even with the O$_3$ outlier omitted from the statistical results), in particular in terms of standard errors, as a direct consequence of the unbalanced treatments of closed-shell molecules and open-shell atoms provided by the model.

For the CCSDT(Q--$n$) models, however, the convergence trends onto the CCSDTQ results are even more pronounced than in Figure \ref{q_n_figure}. As was discussed in Ref. \citenum{quadruples_pert_theory_jcp_2015} in connection with the closed-shell test set, the CCSDT(Q--3) model is practically converged onto the CCSDTQ limit for TAEs, which is encouraging, since the model is still only of moderate cost with respect to the native CCSDTQ model, albeit markedly more costly than, e.g., the $\Lambda$CCSDT(Q) model (cf. the discussion in Section \ref{com_details_section}). Thus, while the model clearly benefits from cancellations of errors, as it does not account for the quadruples relaxation effects included at higher orders in the CCSDT(Q--$n$) series~\cite{quadruples_pert_theory_jcp_2015}, it seems---on the basis of the present study at least---to offer an {\it{a priori}} balanced description of quadruples effects in calculations of relative energy differences for a variety of molecular species, at a reasonable compromise between accuracy and cost, in much the same manner as the CCSD(T--4) model of the CCSD(T--$n$) triples series was found to do for triples effects in the first part of the present series~\cite{open_shell_triples_arxiv_2015}.

%
%
\section{Summary and conclusion}\label{conclusion_section}

In the first part of the present series~\cite{open_shell_triples_arxiv_2015}, the potential of various perturbative CC triples expansions to recover total CCSDT energies for open-shell species was assessed. In this second part, the investigation has been extended to the study of perturbative CC quadruples expansions. In particular, the second- through sixth-order models of the recently proposed CCSDT(Q--$n$) series of quadruples models have been compared to the established CCSDT(Q) and $\Lambda$CCSDT(Q) models by {\textbf{(i)}} evaluating frozen-core/cc-pVDZ total energies against CCSDTQ reference results for two test sets consisting of 18 atoms and small radicals and 17 closed-shell molecules, and {\textbf{(ii)}} comparing TAEs obtained with the same models to CCSDTQ reference results.

In summary, we find that both in terms of the size-intensive recovery of the quadruples contribution to CCSDTQ correlation energy, i.e., the CCSDTQ--CCSDT energy difference, as well as the total deviation from this, the CCSDT(Q--$n$) models all outperform the CCSDT(Q) and $\Lambda$CCSDT(Q) models, with the exception of the lowest-order CCSDT(Q--2) model. The same practical convergence trend onto the target energy (that of the CCSDTQ model) is observed as was previously reported in Ref. \citenum{open_shell_triples_arxiv_2015} for the CCSD(T--$n$) triples models onto the CCSDT energy. While there exist pronounced differences in the description of closed- and open-shell species for the CCSDT(Q) and $\Lambda$CCSDT(Q) models, the CCSDT(Q--$n$) models are found to perform equally well for both test sets, irrespective of the reference determinant used in the open-shell calculations (UHF or ROHF). Upon testing how stable either of the $\Lambda$CCSDT(Q) or CCSDT(Q--$n$) models are by calculating TAEs for all the members of the two test sets, the $\Lambda$CCSDT(Q) results are found to be substantially in error and nowhere near those of the third- and higher-order CCSDT(Q--$n$) models. For the CCSDT(Q--$n$) series, on the other hand, we observe a rapid convergence onto the CCSDTQ reference results. Thus, while we note that more appropriate (e.g., truly isodesmic~\cite{isodesmic_reactions} or even homodesmotic~\cite{homodesmotic_reactions}) reaction schemes exist for the determination of thermochemical quantities, it will not always be feasible to construct such hypothetical reactions (radicals can be difficult in this regard), and in any case, the present comparison in terms of TAEs reports the worst possible errors which, for the third- and higher-order models of the CCSDT(Q--$n$) series, are found to be minor (both relative to CCSDTQ as well as on an absolute scale).

Furthermore, the present study highlights that in order to practically eliminate the remaining uncertainty (that is, make the standard errors diminish for any non-iterative quadruples model), a truly advanced description of quadruples relaxation effects is called for. However, since such a description is only found in the higher-level models of the CCSDT(Q--$n$) series, and since traversing higher up through this hierarchy than, say, the level of the CCSDT(Q--5) and CCSDT(Q--6) models would make the actual savings with respect to the native CCSDTQ calculation negligible, we argue that models such as the CCSDT(Q--3) and CCSDT(Q--4) models will have to suffice as adequate compromises between accuracy and computational cost. Indeed, if even larger errors than those encountered at these levels are acceptable, then, for instance, the $\Lambda$CCSDT(Q) model is an ideal candidate, giving reasonable results (e.g., better than the CCSDT(Q) model) at a relatively low cost. However, our findings lead us to stress that this model, established as it might be, is clearly not balanced in the description of closed- and open-shell species, in much the same way as was previously observed for the CCSD(T) triples model in the first part of this series. Thus, despite its unambiguous formulation for both RHF, UHF, and ROHF references, the $\Lambda$CCSDT(Q) model might occasionally provide answers that are significantly in error with respect to higher-level CC models. 

Finally, one might argue that the errors affiliated with the choice of basis set in the present study (cc-pVDZ) are bound to be in excess of the error bars for, e.g., the TAEs in Figure \ref{ae_figure}. However, accurate calculations of quadruples effects in anything larger than basis sets of double-$\zeta$ quality have unfortunately not yet become standard protocol, and we note that CCSDTQ calculations of this very type (i.e., limited to the cc-pVDZ basis set) are integrated parts of various composite thermochemical models, which is one of the premier domains in which an account of quadruples effects is mandatory. For this reason, and in combination with the minimal basis set dependence previously observed for the CCSDT(Q--$n$) models in Ref. \citenum{quadruples_pert_theory_jcp_2015}, we argue that the higher-order CCSDT(Q--$n$) results reported herein---in particular those of the CCSDT(Q--3) model for the closed- as well as the open-shell test set---are indeed highly encouraging.

%
%
\section*{Acknowledgments}

J. J. E. and P. J. acknowledge support from the European Research Council under the European Union's Seventh Framework Programme (FP/2007-2013)/ERC Grant Agreement No. 291371. J. G. acknowledges financial support from the Deutsche Forschungsgemeinschaft (DFG GA 370/5-1), and D. A. M. acknowledges support from the US National Science Foundation (NSF) under grant number ACI-1148125/1340293.

\newpage

\providecommand*\mcitethebibliography{\thebibliography}
\csname @ifundefined\endcsname{endmcitethebibliography}
  {\let\endmcitethebibliography\endthebibliography}{}

\end{document}